\documentclass[aip,apl,reprint,showpacs,superscriptaddress]{revtex4-1}
\usepackage[T1]{fontenc}
\usepackage{amsfonts,amsmath,amssymb,amsbsy}
\usepackage{graphicx,color}
\usepackage{times}
\definecolor{yblue}{rgb}{0.06, 0.30, 0.57}
\usepackage[colorlinks=true, linkcolor=yblue, citecolor=yblue, urlcolor=yblue]{hyperref}
\let\mathbf=\boldsymbol
\def\emph#1{\textcolor{red}{#1}}
\def\emph#1{\textcolor{black}{#1}}
\begin{document}
%%%%%%%%%%%%%%%%%%%%%%%%%%%%%%%%%%%%%%%%%%%%%%%%%%%%%%%%%%%%

\title{Controllable transport of a skyrmion in a ferromagnetic narrow channel with voltage-controlled magnetic anisotropy}

\author{Junlin Wang}
\thanks{These authors contributed equally to this work.}
\affiliation{School of Science and Engineering, The Chinese University of Hong Kong, Shenzhen 518172, China}
\affiliation{School of Electronic Science and Engineering, Nanjing University, Nanjing 210093, China}
\affiliation{Department of Electronic Engineering, University of York, York, YO10 5DD, United Kingdom}

\author{Jing Xia}
\thanks{These authors contributed equally to this work.}
\affiliation{School of Science and Engineering, The Chinese University of Hong Kong, Shenzhen 518172, China}

\author{Xichao Zhang}
\affiliation{School of Science and Engineering, The Chinese University of Hong Kong, Shenzhen 518172, China}
\affiliation{School of Electronic Science and Engineering, Nanjing University, Nanjing 210093, China}

\author{G. P. Zhao}
\affiliation{College of Physics and Electronic Engineering, Sichuan Normal University, Chengdu 610068, China}

\author{Jing Wu}
\affiliation{Department of Physics, University of York, York, YO10 5DD, United Kingdom}

\author{Yongbing Xu}
\email[E-mail:~]{yongbing.xu@york.ac.uk}
\affiliation{School of Electronic Science and Engineering, Nanjing University, Nanjing 210093, China}
\affiliation{Department of Electronic Engineering, University of York, York, YO10 5DD, United Kingdom}

\author{\\ Zhigang~Zou}
\affiliation{Eco-materials and Renewable Energy Research Center (ERERC), National Laboratory of Solid State Microstructures, Department of Physics, Nanjing University, Nanjing 210093, China}

\author{Yan Zhou}
\email[E-mail:~]{zhouyan@cuhk.edu.cn}
\affiliation{School of Science and Engineering, The Chinese University of Hong Kong, Shenzhen 518172, China}

%-%-%-%-%-%-%-%-%-%-%-%-%-%-%-%-%-%-%-%-%-%-%-%-%-%-%-%-%-%-%
\begin{abstract}
Magnetic skyrmions have potential applications in next-generation spintronics devices with ultralow energy consumption. In this work, the current-driven skyrmion motion in a narrow ferromagnetic nanotrack with voltage-controlled magnetic anisotropy (VCMA) is studied numerically. By utilizing the VCMA effect, the transport of skyrmion can be unidirectional in the nanotrack, leading to a one-way information channel. The trajectory of the skyrmion can also be modulated by periodically located VCMA gates, which protects the skyrmion from destruction by touching the track edge. In addition, the location of the skyrmion can be controlled by adjusting the driving pulse length in the presence of the VCMA effect. Our results provide guidelines for practical realization of the skyrmion-based information channel, diode, and racetrack memory.
\end{abstract}
%-%-%-%-%-%-%-%-%-%-%-%-%-%-%-%-%-%-%-%-%-%-%-%-%-%-%-%-%-%-%

\date{12 September 2017}
%\keywords{magnetic skyrmions, spintronics, micromagnetics}
\pacs{75.60.Ch, 75.70.Kw, 75.78.Cd, 12.39.Dc}
% Domain walls and domain structure:                  75.60.Ch
% Domain structure (magnetic bubbles and vortices):   75.70.Kw
% Micromagnetic simulations:                          75.78.Cd
% Skyrmions:                                          12.39.Dc

\maketitle

%-%-%-%-%-%-%-%-%-%-%-%-%-%-%-%-%-%-%-%-%-%-%-%-%-%-%-%-%-%-%
%\section{Introduction}
%\label{se:Introduction}
%-%-%-%-%-%-%-%-%-%-%-%-%-%-%-%-%-%-%-%-%-%-%-%-%-%-%-%-%-%-%

Magnetic skyrmions are nanoscale particle-like topological configurations, which have been found in certain magnetic bulks, films and nanowire~\cite{muhlbauer2009skyrmion, pfleiderer2010skyrmion, munzer2010skyrmion, yu2011near,heinze2011spontaneous,seki2012observation,do2009skyrmions,chen2015room}. The skyrmion is stabilized by delicate competitions among the ferromagnetic exchange coupling, perpendicular magnetic anisotropy (PMA) and Dzyaloshinskii-Moriya interaction (DMI) in magnetic systems~\cite{nagaosa2013topological,fert2013skyrmions,yu2014biskyrmion,du2014highly,du2015edge,jiang2016direct,jiang2015blowing}. Magnetic skyrmions are expected to be used as information carriers in the next-generation spintronic devices due to their low-power consumption and small sizes~\cite{nii2015uniaxial,zhang2015magnetic,wang2017magnetic,zhang2015magnetic2,zhang2015skyrmion,liu2017chopping,nakatani2016electric}.
In this \textit{Letter}, we report the dynamics of a skyrmion in a narrow ferromagnetic nanotrack channel with voltage-controlled perpendicular magnetic anisotropy, which can be used to build the skyrmion diode and ratchet memory~\cite{franken2012shift,sanchez2017analysis}. The pinning and depinning of the magnetic skyrmion in the nanotrack through the voltage-controlled magnetic anisotropy (VCMA) are investigated. This work will be useful for the design and development of the skyrmion transport channel, which is a building block for any future skyrmion-based information devices.

%%%%%%%%%%%%%%%%%%%%%%%%%%%%%%%%%%%%%%%%%%%%%%%%%%%%%%%%%%%%
\begin{figure}[t]
\centerline{\includegraphics[width=0.50\textwidth]{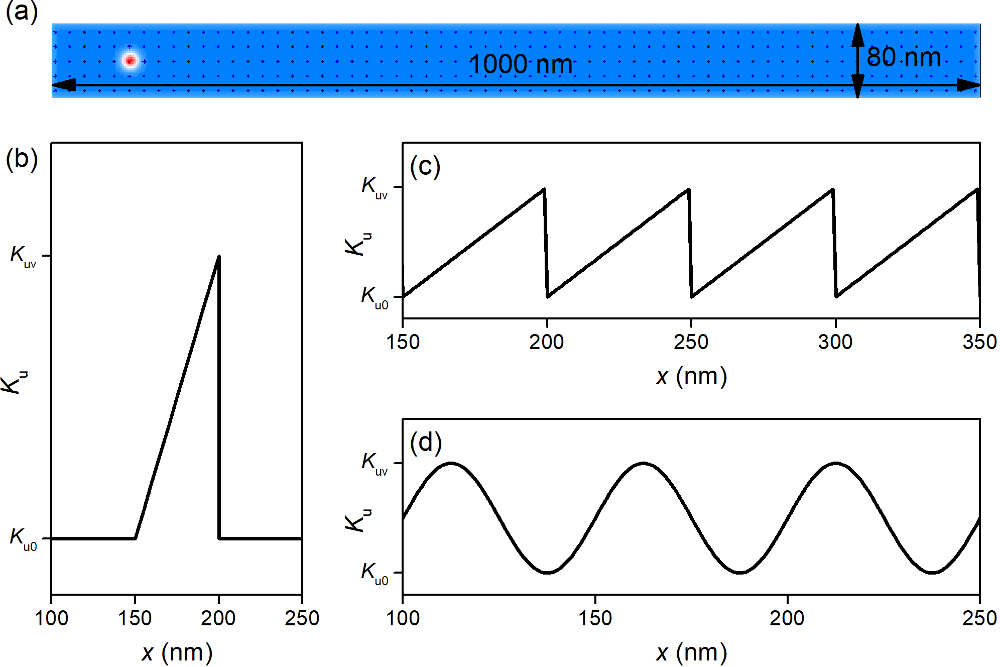}}
\caption{%
(a) A Schematic of the magnetic nanotrack where a magnetic skyrmion is initially placed. The out-of-plane magnetization component is represented by the red ($-z$)-white ($0$)-blue ($+z$) color scale. (b) A linear anisotropy profile. (c) A periodical repetition of a linear anisotropy profile with a period $w$. (b) Sinusoidal function of $x$ with a period $w$.
}
\label{FIG1}
\end{figure}
%%%%%%%%%%%%%%%%%%%%%%%%%%%%%%%%%%%%%%%%%%%%%%%%%%%%%%%%%%%%

%%%%%%%%%%%%%%%%%%%%%%%%%%%%%%%%%%%%%%%%%%%%%%%%%%%%%%%%%%%%
\begin{figure}[t]
\centerline{\includegraphics[width=0.50\textwidth]{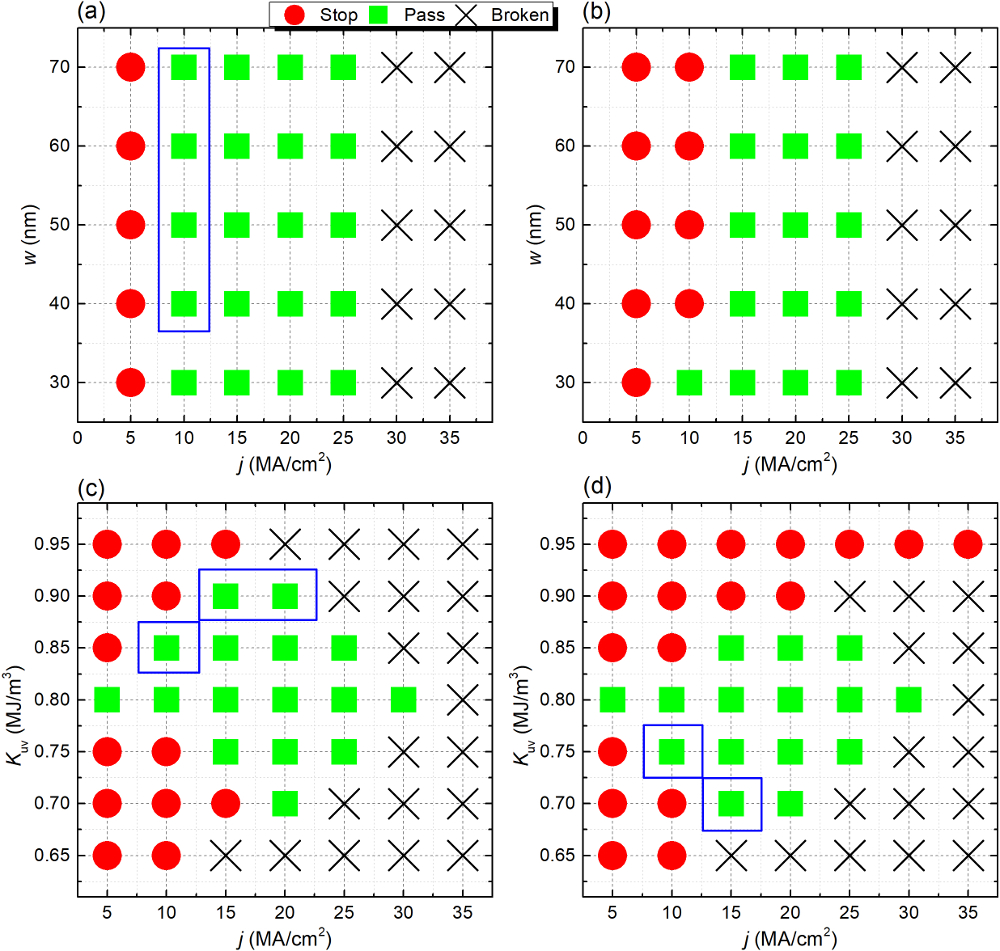}}
\caption{%
The pinning/depinning states of an isolate skyrmion driven by the current in a magnetic track. (a), (b) The pinning/depinning states of a skyrmion at various width $w$ and driving current $j$ along $+x$ and $-x$ axis for $K_\text{uv} = 0.85$~MJ/m$^3$, respectively. (c), (d) The pinning/depinning states of a skyrmion at various $K_\text{uv}$ and $j$ along $+x$ and $-x$ axis for the fixed $w=50$ nm, respectively. The solid circle means the skyrmion is not able to pass the well or barrier, the solid square means the skyrmion can pass the well or barrier and the cross means the skyrmion is destroyed. 
}
\label{FIG2}
\end{figure}
%%%%%%%%%%%%%%%%%%%%%%%%%%%%%%%%%%%%%%%%%%%%%%%%%%%%%%%%%%%%

%%%%%%%%%%%%%%%%%%%%%%%%%%%%%%%%%%%%%%%%%%%%%%%%%%%%%%%%%%%%
\begin{figure}[t]
\centerline{\includegraphics[width=0.50\textwidth]{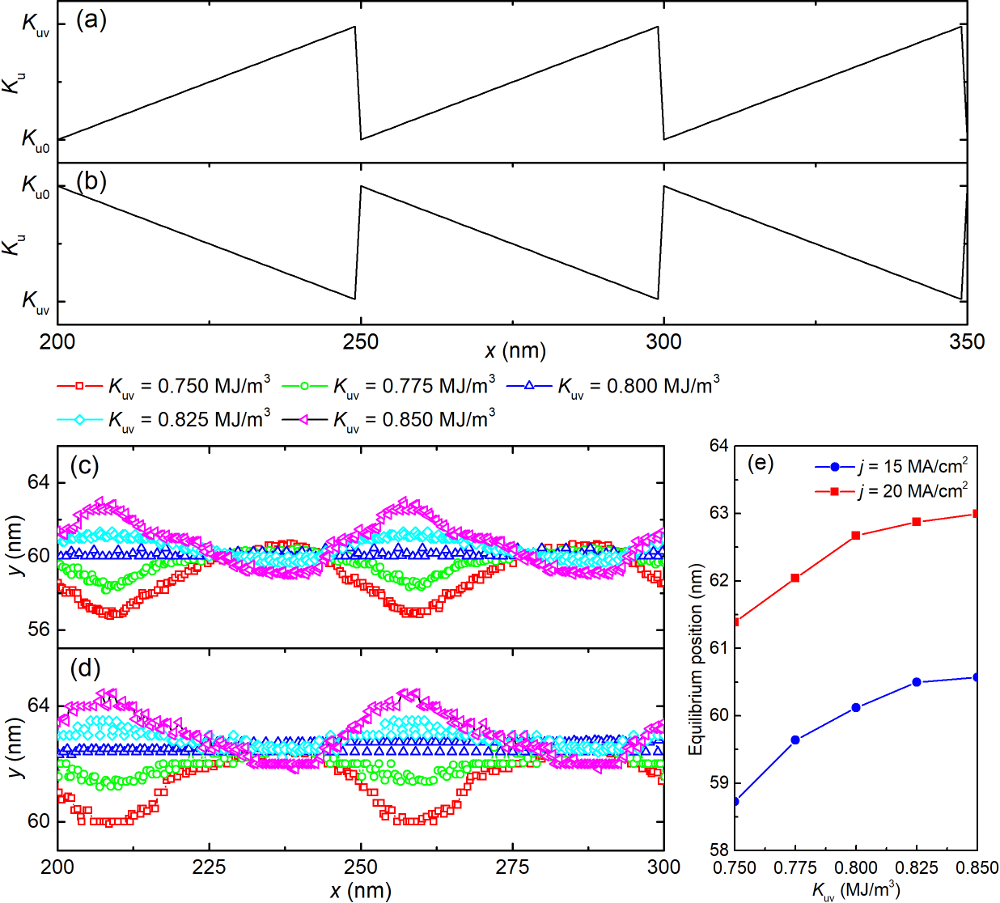}}
\caption{%
(a) The wedge-shaped profile of $K_\text{u}$ for $K_\text{uv}>K_\text{u0}$. (b) The wedge-shaped profile of $K_\text{u}$ for $K_\text{uv}<K_\text{u0}$. (c) The trajectories of the skyrmion in the nanotrack with various $K_\text{uv}$ for $j=15$~MA/cm$^2$. (d) The trajectories of the skyrmion in the nanotrack with various $K_\text{uv}$ for $j=20$~MA/cm$^2$. (e) The equilibrium position of the skyrmion in the $y$ direction for (b) and (c). The spin current is applied along $+x$ axis.
}
\label{FIG3}
\end{figure}
%%%%%%%%%%%%%%%%%%%%%%%%%%%%%%%%%%%%%%%%%%%%%%%%%%%%%%%%%%%%

%%%%%%%%%%%%%%%%%%%%%%%%%%%%%%%%%%%%%%%%%%%%%%%%%%%%%%%%%%%%
\begin{figure}[t]
\centerline{\includegraphics[width=0.50\textwidth]{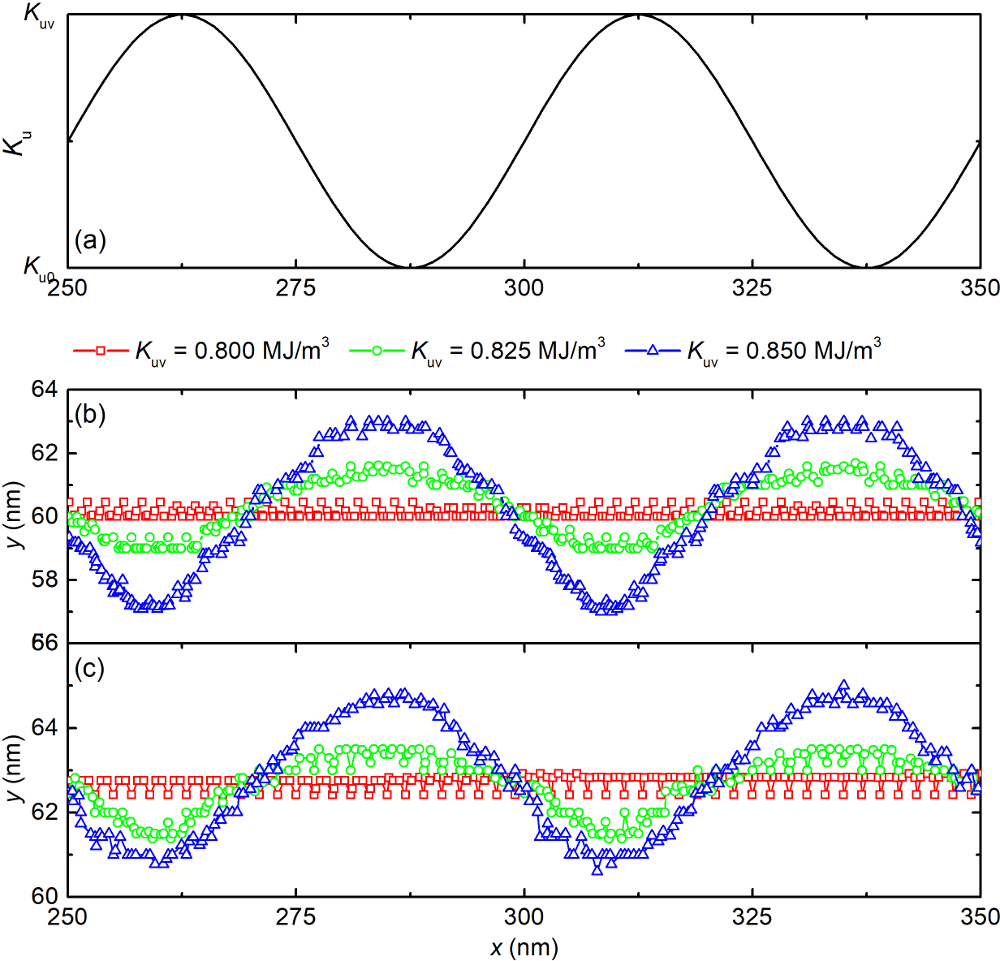}}
\caption{%
(a) The profile of $K_\text{u}$ as a sinusoidal function of $x$. (b) The trajectories of the skyrmion in the nanotrack with various $K_\text{uv}$ for $j=15$~MA/cm$^2$. (c) The trajectories of the skyrmion in the nanotrack with various $K_\text{u}$ for $j=20$~MA/cm$^2$. The spin current is applied along $+x$ axis.
}
\label{FIG4}
\end{figure}
%%%%%%%%%%%%%%%%%%%%%%%%%%%%%%%%%%%%%%%%%%%%%%%%%%%%%%%%%%%%

%%%%%%%%%%%%%%%%%%%%%%%%%%%%%%%%%%%%%%%%%%%%%%%%%%%%%%%%%%%%
\begin{figure}[t]
\centerline{\includegraphics[width=0.50\textwidth]{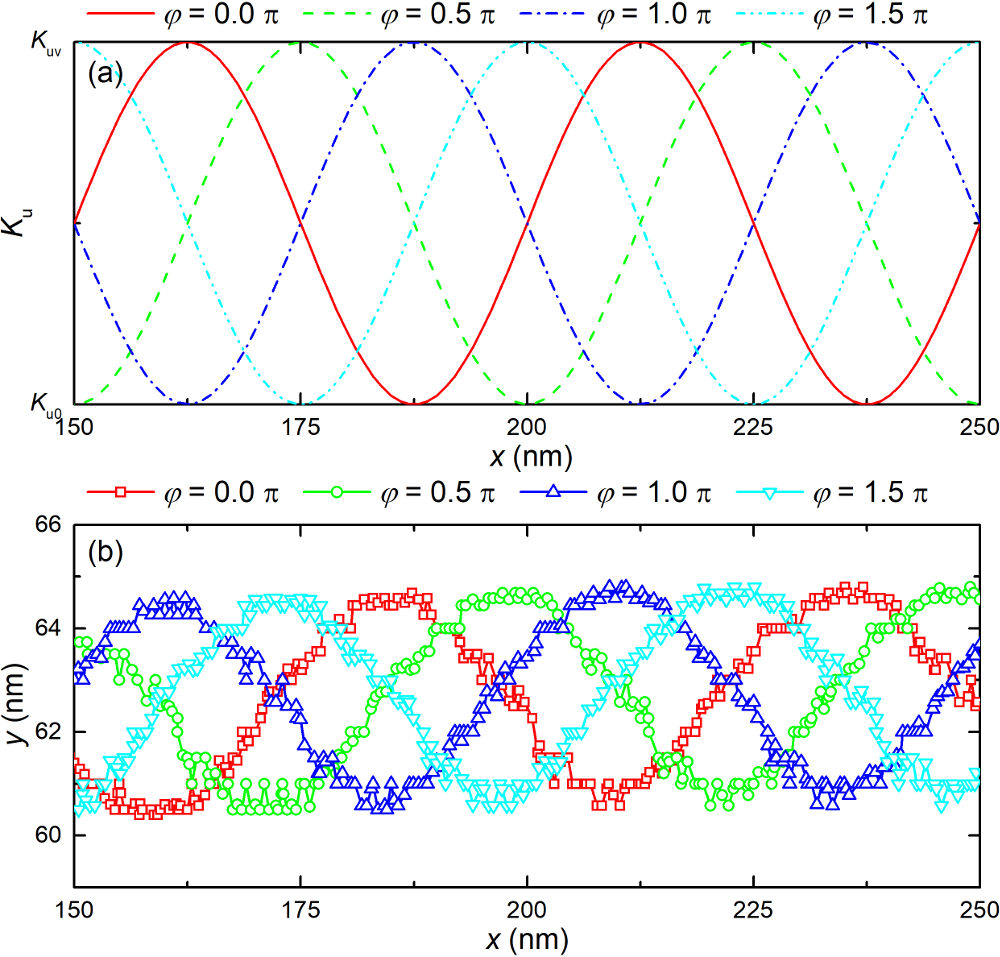}}
\caption{%
(a) The profile of $K_\text{u}$ and (b) the corresponding trajectories of the skyrmion in the nanotrack with $\varphi = 0, 0.5\pi, 1.0\pi, 1.5\pi$. The driving current density is $20$~MA/cm$^2$ applied along $+x$ axis and $K_\text{uv}=0.850$~MJ/m$^3$.
}
\label{FIG5}
\end{figure}
%%%%%%%%%%%%%%%%%%%%%%%%%%%%%%%%%%%%%%%%%%%%%%%%%%%%%%%%%%%%

%%%%%%%%%%%%%%%%%%%%%%%%%%%%%%%%%%%%%%%%%%%%%%%%%%%%%%%%%%%%
\begin{figure}[t]
\centerline{\includegraphics[width=0.50\textwidth]{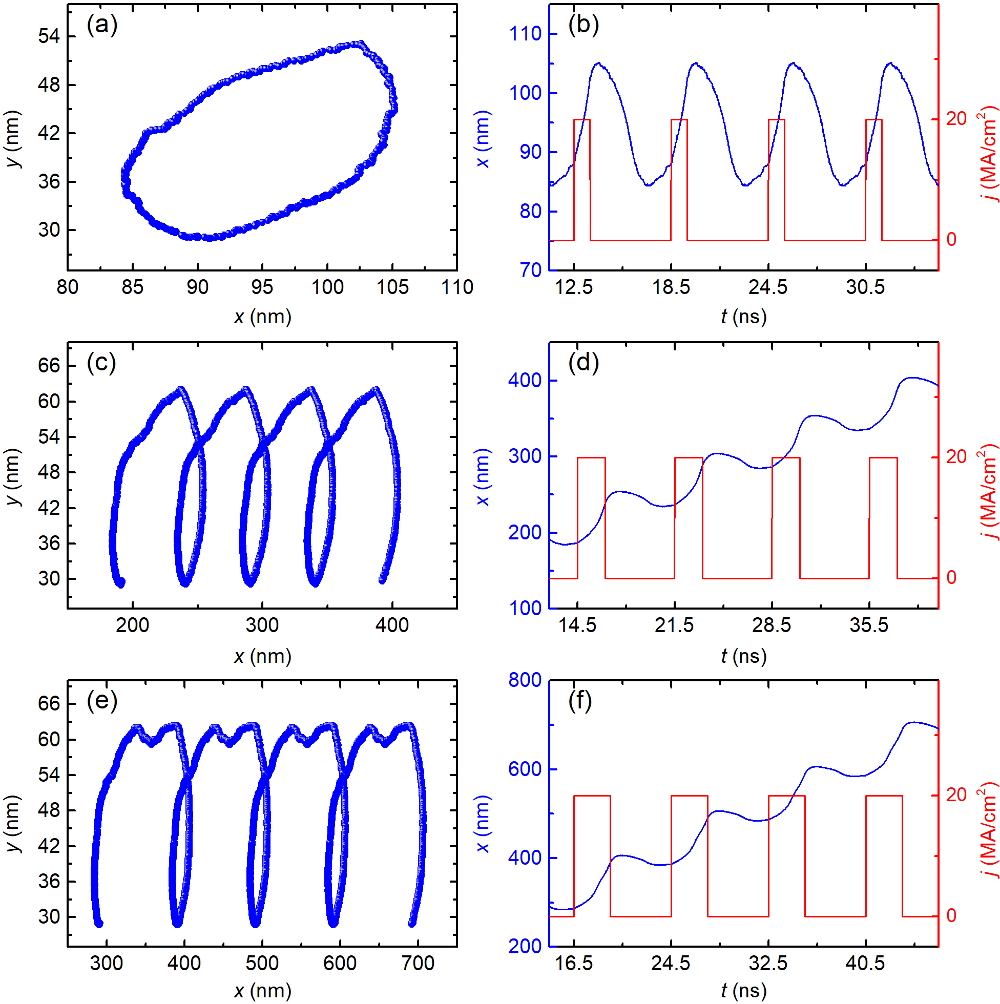}}
\caption{%
The skyrmion motion driven by the current pulse in the nanotrack with the wedge-shaped $K_\text{u}$ with $K_\text{uv}=0.75$~MJ/m$^3$. The left panel shows the trajectories of the skyrmion. The right panel shows the $x$ position of the skyrmion and the current density as functions of time $t$. For one period of the current pulse , $t_e$ is the pulse time and $t_r$ is the relax time without applying current. $t_r=5$~ns in the simulations. (a), (b) $t_e=1$~ns. (c), (d) $t_e=2$~ns. (c), (d) $t_e=3$~ns.
}
\label{FIG6}
\end{figure}
%%%%%%%%%%%%%%%%%%%%%%%%%%%%%%%%%%%%%%%%%%%%%%%%%%%%%%%%%%%%

%%%%%%%%%%%%%%%%%%%%%%%%%%%%%%%%%%%%%%%%%%%%%%%%%%%%%%%%%%%%
\begin{figure}[t]
\centerline{\includegraphics[width=0.50\textwidth]{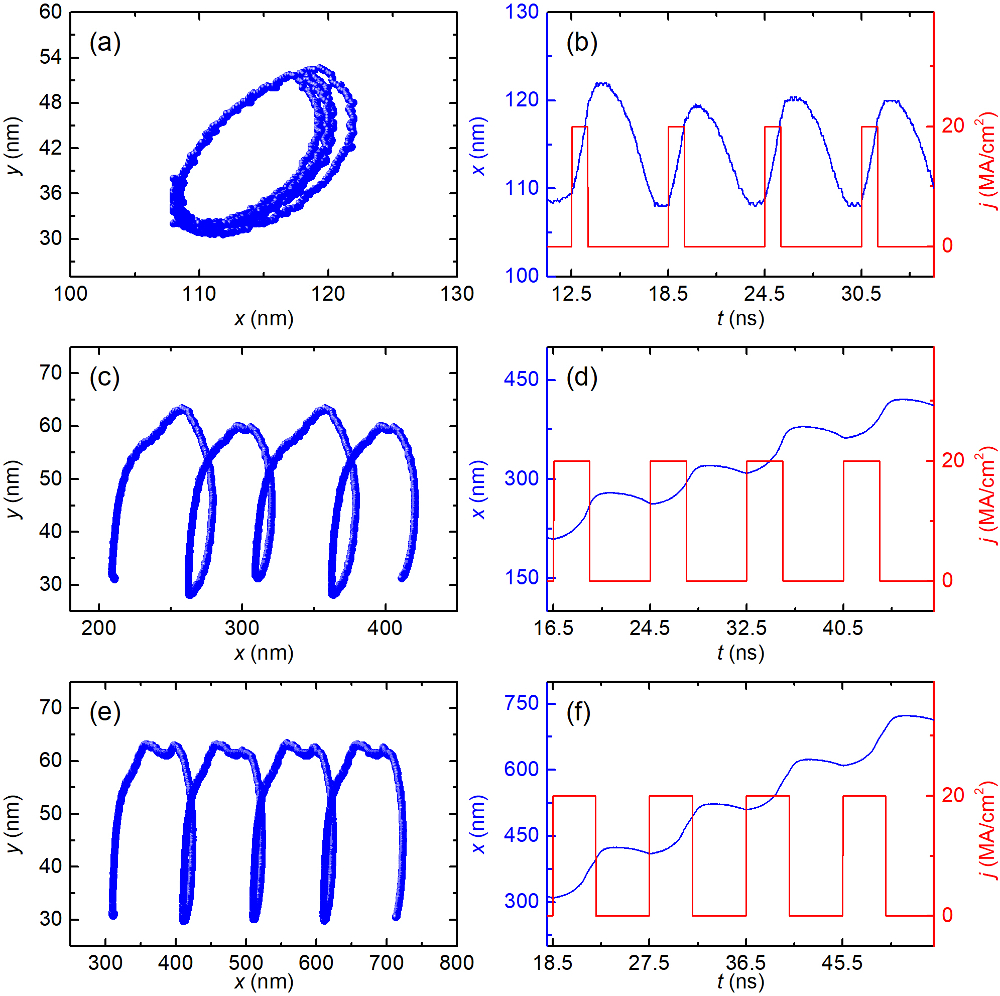}}
\caption{%
The skyrmion motion driven by the current pulse for the wedge-shaped $K_\text{u}$ with $K_\text{uv}=0.85$~MJ/m$^3$. The left panel shows the trajectories of the skyrmion. The right panel shows the $x$ position of the skyrmion and the current as functions of time $t$. $t_r=5$~ns in the simulations. (a), (b) $t_e=1$~ns. (c), (d) $t_e=3$~ns. (c), (d) $t_e=4$~ns.
}
\label{FIG7}
\end{figure}
%%%%%%%%%%%%%%%%%%%%%%%%%%%%%%%%%%%%%%%%%%%%%%%%%%%%%%%%%%%%

%%%%%%%%%%%%%%%%%%%%%%%%%%%%%%%%%%%%%%%%%%%%%%%%%%%%%%%%%%%%
\begin{figure}[t]
\centerline{\includegraphics[width=0.50\textwidth]{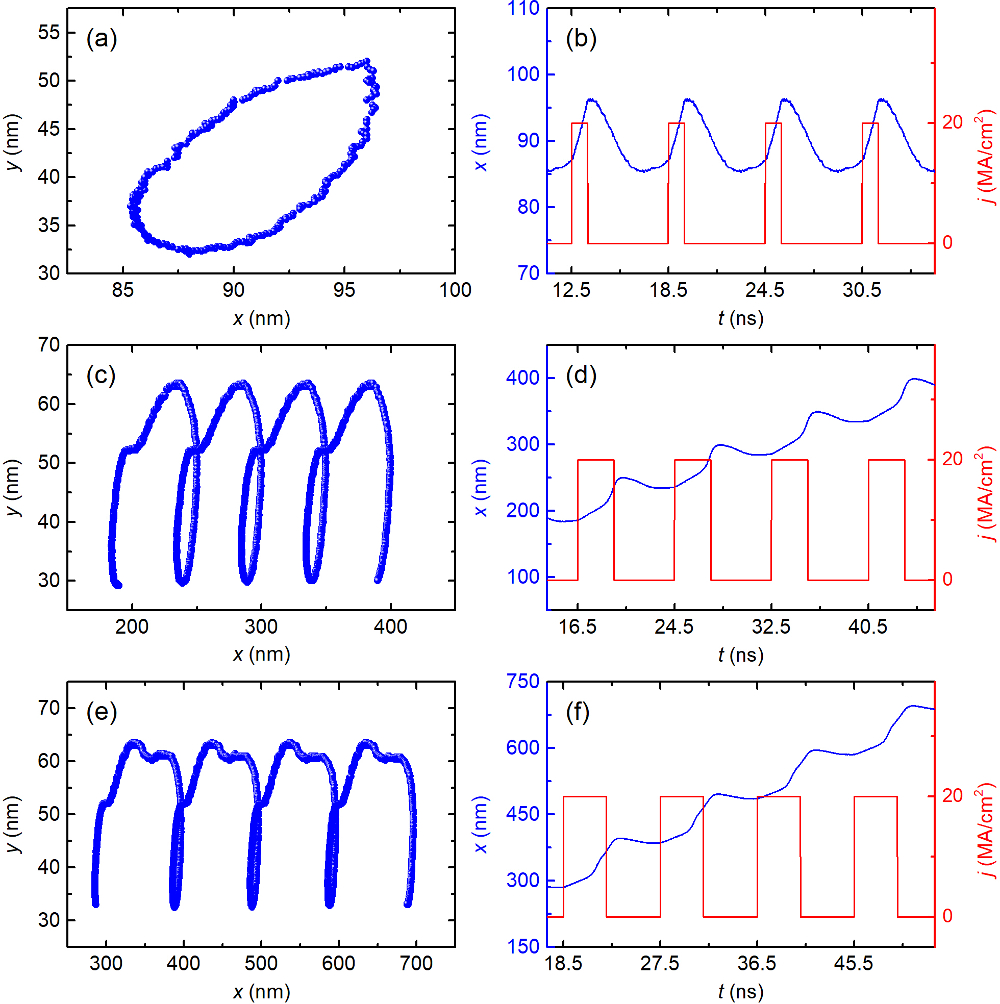}}
\caption{%
The skyrmion motion driven by the current pulse for the sinusoidal $K_\text{u}$ with $K_\text{uv}=0.85$~MJ/m$^3$ and $\varphi=0$. The left panel shows the trajectories of the skyrmion. The right panel shows the $x$ position of the skyrmion and the current as functions of time. $t_r=5$~nm in the simulations. (a), (b) $t_e=1$~ns. (c), (d) $t_e=3$~ns. (c), (d) $t_e=4$~ns.
}
\label{FIG8}
\end{figure}
%%%%%%%%%%%%%%%%%%%%%%%%%%%%%%%%%%%%%%%%%%%%%%%%%%%%%%%%%%%%

%-%-%-%-%-%-%-%-%-%-%-%-%-%-%-%-%-%-%-%-%-%-%-%-%-%-%-%-%-%-%
%\section{Methods}
%\label{se:Methods}
%-%-%-%-%-%-%-%-%-%-%-%-%-%-%-%-%-%-%-%-%-%-%-%-%-%-%-%-%-%-%

\emph{The} simulation model is an ultrathin ferromagnetic nanotrack, $1000~\rm{nm}\times 80~\rm{nm}\times 0.4~\rm{nm}$, as shown in Fig.~\ref{FIG1}a. The micromagnetic simulations are performed with the Object Oriented MicroMagnetic Framework (OOMMF)~\cite{donahue2010oommf}. The dynamic of magnetization are described by Landau-Lifshitz-Gilbert LLG (LLG) equation, written as
\begin{equation}
\frac{d\boldsymbol{m}}{dt}=-\gamma_{0}\boldsymbol{m}\times\boldsymbol{h}_{\rm{eff}}+\alpha(\boldsymbol{m}\times\frac{d\boldsymbol{m}}{dt})-u\boldsymbol{m}\times(\boldsymbol{m}\times\boldsymbol{p}),
\label{eq:LLG}
\end{equation}
where $\boldsymbol{m}$ is the reduced magnetization $\frac{\boldsymbol{M}}{M_\text{S}}$, $M_\text{S}$ is the saturation magnetization. $\gamma_{0}$ is the gyromagnetic ratio and $\alpha$ is the damping coefficient. $\boldsymbol{h}_{\text{eff}}$ is the effective field including the contributions of Heisenberg exchange, Dzyaloshinskii-Moriya interaction (DMI), magnetic anisotropy and demagnetization field. The \textit{u} can be defined as $\frac{\gamma_{0}\hbar jP}{2de\mu_{0}M_{\text{S}}}$, $\hbar$ is the reduced Plank constant, $j$ is the current density, $P=0.08$ is the spin Hall angle, $a$ is the atomic lattice constant, $e$ is the electron charge, $\mu_{0}$ is the vacuum permeability constant, $d$ is the thickness of the magnetic nanotrack\cite{woo2017spin}. $\boldsymbol{p}$ is the direction of the spin polarization which is equal to $-\hat{y}$. The model is discretized into tetragonal volume elements with the size of $2~\text{nm}\times 2~\text{nm}\times 0.4~\text{nm}$. The parameters for the micromagnetic simulation are adopted from Ref.~\onlinecite{wiesendanger2016nanoscale}: the saturation magnetization $M_{\text{S}}=580$~kA/m, the damping coefficient $\alpha=0.3$, the DMI constant $D=3$~mJ/m$^2$, and the exchange constant $A=15$~pJ/m. In the simulation, the profile of the voltage-controlled magnetic anisotropy (VCMA) in the nanotrack are shown in Figs.~\ref{FIG1}b-d. For the simulation of the pinning/depinning states of the skyrmion, the PMA profile is shown in Fig.~\ref{FIG1}b. VCMA linearly varies from $K_{\text{u0}}$ to $K_{\text{uv}}$ and $K_\text{u0} = 0.8$~MJ/m$^3$. For the simulation of the motion of skyrmion, two types of VCMA profile are considered, period wedge-shape and sinusoidal functions, as shown in Figs.~\ref{FIG1}c and \ref{FIG1}d respectively. The function for the period wedge-shape profile is given as:
\begin{equation}
K_{\text{u}}(x) = K_{\text{u0}} + \dfrac{K_{\text{uv}} - K_{\text{u0}}}{w}x,  
\label{eq:Wedge-shape}
\end{equation}
\begin{equation}
K_{\text{u}}(x) = {K}_{\text{u0}} + \dfrac{K_{\text{uv}} - K_{\text{u0}}}{2}(1+\sin\left(2\pi x/w-\varphi\right)),
\label{eq:Sinusoidal-function}
\end{equation}
where $w$ is the period length $w$, $\varphi$ is the phase, and $x$ is the longitudinal coordinate.
The period wedge-shape is given in the Eq.~\ref{eq:Wedge-shape} and the sinusoidal function is given in the Eq.~\ref{eq:Sinusoidal-function}. The linear anisotropy profile and the sinusoidal function profile are given in the Figs.~\ref{FIG1}b and \ref{FIG1}c.

%-%-%-%-%-%-%-%-%-%-%-%-%-%-%-%-%-%-%-%-%-%-%-%-%-%-%-%-%-%-%
%\section{Results}
%\label{se:Results}
%-%-%-%-%-%-%-%-%-%-%-%-%-%-%-%-%-%-%-%-%-%-%-%-%-%-%-%-%-%-%

Fig.~\ref{FIG2} shows the pinning/depinning states of isolate skyrmion driven by the spin current in a nanotrack with the PMA profile shown in Fig.~\ref{FIG1}b. Figures~\ref{FIG2}a and \ref{FIG2}b show the effect of the width and the current density on the pinning/depinning states. Initially, the relaxed skyrmion is located at the left side of the VCMA region when the spin current is applied along $+x$ axis. The skyrmion is not able to pass the VCMA region when the current density is smaller than $10$~MA/cm$^2$ and pass the VCMA region when $25~\text{MA/cm}^2<j<30~\text{MA/cm}^2$. The skyrmion will be destroyed when the current is larger than $30$~MA/cm$^2$. When the spin current is applied along $-x$ axis, the skyrmion is located at the right side of the VCMA region. Most states are the same to the corresponding results in Fig.~\ref{FIG1}a, except for the case of $j=10~\text{MA/cm}^2$. For $j=10~\text{MA/cm}^2$ and $w>30~\text{nm}$, the skyrmion can pass the VCMA region when the current is applied along $+x$ axis while it can not pass when the current is applied along $-x$ axis. It means that the skyrmion can pass only in one direction, $+x$ axis. The motion of skyrmion is unidirectional. The parameters corresponding to the unidirectional pass along $+x$ axis are marked with blue box in Fig.~\ref{FIG2}a. Figures~\ref{FIG2}c and \ref{FIG2}d show the effect of the VCMA and the current density on the pinning/depinning states. The results shows that the states is sensitive to the VCMA. The unidirectional behaviors also can be found. The parameters for the unidirectional pass along $+x$ axis are marked with blue box in Fig.~\ref{FIG2}c and these for the unidirectional pass along $-x$ axis are marked with blue box in Fig.~\ref{FIG2}d. The unidirectional behaviors shows that the voltage gate can be used to built skyrmion diode.

The skyrmion motion driven by the spin current in a magnetic nanotrack with the spatially dependence of VCMA is simulated. The VCMA is periodical repetition of a wedge-shape profile, as shown in Figs.~\ref{FIG3}a ($K_\text{uv}>K_\text{u0}$) and \ref{FIG3}b ($K_\text{uv}>K_\text{u0}$). Initially, the relaxed skyrmion is located at $x=86$~nm and $y=40$~nm. The trajectories of the skyrmion driven by the spin current ($j=15~\text{MA/cm}^2$) in the nanotrack with various $K_\text{uv}$ are shown in Fig.~\ref{FIG3}c. For $K_\text{uv}=0.800~\text{MJ/m}^3$, a uniform perpendicular magnetic anisotropy in the nanotrack, the skyrmion shows a transverse motion towards to the upper edge resulted by the transverse force due to skyrmion Hall effect firstly~\cite{jiang2016direct}. When the transverse force and edge-skyrmion repulsive force are balanced, the skyrmion moves straightly~\cite{du2015edge,nii2015uniaxial,jiang2015blowing}. It can be seen that the skyrmion moves straightly at $y=60$~nm finally. For $K_\text{uv}=0.850~\text{MJ/m}^3$, the skyrmion moves in a periodical wavy trajectory with an equilibrium position at $y=60.6$~nm. Similar behaviors of the skyrmion are found when $K_\text{uv}=0.750~\text{MJ/m}^3,0.775~\text{MJ/m}^3$, and $0.825~\text{MJ/m}^3$. It can be found that the equilibrium position increases with increasing $K_\text{uv}$, which is shown in Fig.~\ref{FIG3}e. Periodical wavy trajectories and similar dependence of the equilibrium position on $K_\text{uv}$ can be also found in the case of $j=20~\text{MA/cm}^2$, as shown in Fig.~\ref{FIG3}d. The equilibrium positions of the periodical wavy trajectories is larger compared to the case of $j=15~\text{MA/cm}^2$.

Figure~\ref{FIG4} shows the trajectories of the skyrmion in a nanotrack with sinusoidal dependence of $K_\text{u}$ on the position $x$. The profile of $K_\text{u}$ is shown in Fig.~\ref{FIG4}a. $K_\text{u0}$ is the minimum and $K_\text{uv}$ is the maximum. It can be found from Fig.~\ref{FIG4}b that the skyrmion moves in a sinusoidal trajectory when $K_\text{uv}\neq 0.8~\text{MJ/m}^3$. Differently from the case of the wedge-shaped profile of $K_\text{u}$, the equilibrium positions in $y$ direction for various $K_\text{uv}$ are almost the same, $y=60$~nm. When the current density increases to $j=20~\text{MA/cm}^2$, similar results can be found. Further, the effect of the phase also has been simulated and the results are shown in Fig.~\ref{FIG5}.

The motion of the magnetic skyrmion in the nanotrack with VCMA driven by the current pulse also be simulated. The initial position the skyrmion is  $x =~86$ nm which is the middle of a voltage gate. Figure~\ref{FIG6} shows the motion of the skyrmion in the nanotrack with a periodical wedge-shaped profile with $K_\text{uv}=0.750~\text{MJ/m}^3$ with the period length $w=50$~nm. The current density of the pulse is $20~\text{MA/cm}^2$. The pulse \emph{is} applied at $t=0.5$~ns. For one period of the current pulse , $t_e$ is the time interval applying the current and $t_r$ is the relax time without applying current. $t_r=5$~ns in the simulations. When $t_e=1$~ns, the skyrmion cannot pass the voltage gate and moves in a circle trajectory as shown in Figs.~\ref{FIG6}a and \ref{FIG6}b. For $t_e=2$~ns, the trajectory of the skyrmion is shown in Fig.~\ref{FIG6}c. The time-dependence of the position in the $x$ direction and the current density are shown in Fig.~\ref{FIG6}d. At $t=14.5$~ns, the skyrmion is located at $x=187$~nm. After applying the pulse, $x=241$~nm at $t=16.5$~ns. Then the applied current is off. The skyrmion further relax to $x=236$~nm before the next pulse. The displacement of skyrmion is $50$~nm after a pulse is applied. For $t_e=3$~ns, Figs.~\ref{FIG6}e and \ref{FIG6}f, one current pulse results in a displacement of $100$~nm.

Figure~\ref{FIG7} shows the results for the case of wedge-shaped $K_{\text{uv}}$ with $K_{\text{uv}} = 0.850~\text{MJ/m}^3$ and the period length $w = 50$~nm. The current density of the pulse is $20~\text{MA/cm}^2$ and the pulse is applied at $t=0.5$~ns with $t_r = 5$~ns. In Figs.~\ref{FIG7}a and \ref{FIG7}b, compared with the state with $K_\text{uv}=0.750~\text{MJ/m}^3$, the skyrmion can more easily pass the voltage gate. This state also has been explained in Fig.~\ref{FIG2}c. The skyrmion passes the first voltage gate and cannot pass the second voltage gate. Then the skyrmion moves in a circle trajectory. When the $t_e=2$~ns and 3~ns, the states is similar as Figs.~\ref{FIG6}c-f. The skyrmion pass one or two voltage gates are shown in Figs.~\ref{FIG7}c-f.
In Figs.~\ref{FIG7}c and \ref{FIG7}d, for $t_e=2$~ns, the skyrmion is located at $x=212$~nm when $t=16.5$~ns. After applying the pulse, skyrmion moves to $x=277$~nm. When the current is off, the skyrmion further relaxes to $x=263$~nm before the next pulse. The displacement of skyrmion is $50$~nm after a pulse is applied. For $t_e=3$~ns, as shown in Figs.~\ref{FIG7}e and \ref{FIG7}f, one current pulse lead to a displacement of $100$~nm.

Figure~\ref{FIG8} shows the motion of the skyrmion in the nanotrack with a sinusoidal function profile with $K_\text{uv}=0.850~\text{MJ/m}^3$ with the period length $w=50$~nm. The trajectories of skyrmion with $t_e=1$~ns are shown in Figs.~\ref{FIG8}a and \ref{FIG8}b. The skyrmion cannot pass the voltage gate and moves in a circle which is like Figs.~\ref{FIG8}a and \ref{FIG8}b. For $t_e=2$~ns, the trajectory of the skyrmion is shown in Fig.~\ref{FIG8}c. The time-dependence of the position in $x$ direction and the current density are shown in Fig.~\ref{FIG8}d. At $t=16.5$~ns, the skyrmion is located at $x=186$~nm. After applying the pulse, $x=236$~nm at $t=18.5$~ns. Then the applied current is off. The skyrmion further relax to $x=234$~nm before the next pulse. The skyrmion moves with a pulse time $t_e=3$~ns is shown in Figs.~\ref{FIG8}e and \ref{FIG8}f which one current pulse results in a displacement of $100$~nm. From Fig.~\ref{FIG6} to Fig.~\ref{FIG8}, it can be seen that the model with multiple voltage gates can be used to realize high density racetrack memory device.

%-%-%-%-%-%-%-%-%-%-%-%-%-%-%-%-%-%-%-%-%-%-%-%-%-%-%-%-%-%-%
%\section{Conclusions}
%\label{se:Conclusions}
%-%-%-%-%-%-%-%-%-%-%-%-%-%-%-%-%-%-%-%-%-%-%-%-%-%-%-%-%-%-%

In this \textit{Letter}, the skyrmion motion in a ferromagnetic nanotrack with single or multiple VCMA gates is studied. This work shows the trajectory and location of the skyrmion can be controlled by periodically located VCMA gates as well as the driving current pulse. The unidirectional motion of the skyrmion realized by the VCMA effect can be used to build the skyrmion-based one-way information channel and the skyrmion diode. Our results are useful for the design and development of the skyrmion-based spintronic devices.

%-%-%-%-%-%-%-%-%-%-%-%-%-%-%-%-%-%-%-%-%-%-%-%-%-%-%-%-%-%-%
\begin{acknowledgments}
Y.X. acknowledges the support by the State Key Program for Basic Research of China (Grant Nos. 2014CB921101 and 2016YFA0300803), National Natural Science Foundation of China (Grant Nos. 61427812 and 11574137), Jiangsu NSF (No. BK20140054), Jiangsu Shuangchuang Team Program, and the UK EPSRC (EP/G010064/1).
Y.Z. acknowledges the support by the National Natural Science Foundation of China (Grant No. 11574137) and Shenzhen Fundamental Research Fund (Grant No. JCYJ20160331164412545).
G.P.Z. acknowledges the support by the National Natural Science Foundation of China (Grant Nos. 11074179 and 10747007), the Construction Plan for Scientific Research Innovation Teams of Universities in Sichuan (No. 12TD008).
X.Z. was supported by JSPS RONPAKU (Dissertation Ph.D.) Program.
\end{acknowledgments}
%-%-%-%-%-%-%-%-%-%-%-%-%-%-%-%-%-%-%-%-%-%-%-%-%-%-%-%-%-%-%

%-%-%-%-%-%-%-%-%-%-%-%-%-%-%-%-%-%-%-%-%-%-%-%-%-%-%-%-%-%-%

%-%-%-%-%-%-%-%-%-%-%-%-%-%-%-%-%-%-%-%-%-%-%-%-%-%-%-%-%-%-%

%%%%%%%%%%%%%%%%%%%%%%%%%%%%%%%%%%%%%%%%%%%%%%%%%%%%%%%%%%%%
\end{document}